\documentclass[fleqn,usenatbib]{mnras}

\usepackage{newtxtext,newtxmath}
\usepackage[T1]{fontenc}

\DeclareRobustCommand{\VAN}[3]{#2}
\let\VANthebibliography\thebibliography
\def\thebibliography{\DeclareRobustCommand{\VAN}[3]{##3}\VANthebibliography}

\usepackage{graphicx}	
\usepackage{amsmath}	
\usepackage{enumerate}
\usepackage{soul}               
\usepackage{newtxtext,newtxmath}

\title[Anisotropic infall onto the Local Group]{Anisotropic satellite accretion onto the Local Group with HESTIA}

\author[Dupuy et al.]{
Alexandra Dupuy$^{1,2}$\thanks{E-mail: adupuy@kias.re.kr},
Noam I. Libeskind$^{3}$, 
Yehuda Hoffman$^{4}$, 
Helène M. Courtois$^{2}$,
Stefan Gottlöber$^{3}$,
\newauthor{Robert J. J. Grand$^{5,6}$,
Alexander Knebe$^{7,8,9}$,
Jenny G. Sorce$^{10,3}$,
Elmo Tempel$^{11,12}$,
R. Brent Tully$^{13}$,}
\newauthor{Mark Vogelsberger$^{14}$,
Peng Wang$^{15,3}$}
\\
$^{1}$Korea Institute for Advanced Study, 85, Hoegi-ro, Dongdaemun-gu, Seoul 02455, Republic of Korea\\
$^{2}$ University of Lyon, UCB Lyon 1, CNRS/IN2P3, IUF, IP2I Lyon, France\\
$^{3}$Leibniz-Institut f\"ur Astrophysik Potsdam (AIP), An der Sternwarte 16, D-14482 Potsdam, Germany\\
$^{4}$Racah Institute of Physics, Hebrew University, Jerusalem 91904, Israel\\
$^{5}$Instituto de Astrof\'isica de Canarias, Calle Vía L\'actea s/n, E-38205 La Laguna, Tenerife, Spain\\
$^{6}$Departamento de Astrof\'isica, Universidad de La Laguna, Av. del Astrof\'isico Francisco S\'anchez s/n, E-38206, La Laguna, Tenerife, Spain\\
$^{7}$Departamento de Física Teórica, Módulo 15, Facultad de Ciencias, Universidad Autónoma de Madrid, 28049 Madrid, Spain\\
$^{8}$Centro de Investigación Avanzada en Física Fundamental (CIAFF), Facultad de Ciencias, Universidad Autónoma de Madrid, 28049 Madrid, Spain \\
$^{9}$International Centre for Radio Astronomy Research, University of Western Australia, 35 Stirling Highway, Crawley, Western Australia 6009, Australia \\
$^{10}$Universit\'e Paris-Saclay, CNRS, Institut d'Astrophysique Spatiale, 91405, Orsay, France\\
$^{11}$ Tartu Observatory, University of Tartu, Observatooriumi 1, 61602 T\~oravere, Estonia\\
$^{12}$ Estonian Academy of Sciences, 10130 Kohtu 6, Tallinn, Estonia\\
$^{13}$ Institute for Astronomy (IFA), University of Hawaii, 2680 Woodlawn Drive, HI 96822, USA\\
$^{14}$ Department of Physics, Kavli Institute for Astrophysics and Space Research, Massachusetts Institute of Technology, Cambridge, MA 02139, USA\\
$^{15}$ Shanghai Astronomical Observatories, Shanghai 200030, China 
}

\date{Accepted 2022 August 26. Received 2022 August 23; in original form 2022 June 2}

\pubyear{2022}

\begin{document}
\label{firstpage}
\pagerange{\pageref{firstpage}--\pageref{lastpage}}
\maketitle

\begin{abstract}
How the cosmic web feeds halos, and fuels galaxy formation is an open question with wide implications. This study explores the mass assembly in the Local Group within the context of the local cosmography by employing simulations whose initial conditions have been constrained to reproduce the local environment. The goal of this study is to inspect whether the direction of accretion of satellites on to the Milky Way and Andromeda galaxies, is related to the cosmic web. The analysis considers the three high-resolution simulations available in the HESTIA simulation suite, as well as the derived velocity shear and tidal tensors. We notice two eras in the Local Group accretion history, delimited by an epoch around $z \approx 0.7$. We also find that satellites can travel up to $\sim 4$ Mpc, relative to their parent halo before crossing its viral radius $R_{200}$. Finally, we observe a strong alignment of the infall direction with the axis of slowest collapse $\vec{e_3}$ of both tidal and shear tensors, implying satellites of the Local Group originated from one particular region of the cosmic web and were channeled towards us via the process of accretion.This alignment is dominated by the satellites that enter during the early infall era, i.e $z>0.7$.
\end{abstract}



\section{Introduction}

The cosmological principle states that the Universe is isotropic and homogeneous on large enough scales. This implies that there are no special places nor special directions in the Universe. Nevertheless, this is not the case on smaller scales, where we observe structures suggesting the existence of preferred directions, such as the Supergalactic plane, recognised by \cite{1956VA......2.1584D, 1975ApJ...202..610D}. 

The notion of {\it cosmic web} \citep{1996Natur.380..603B} describes the anisotropic mass assembly of matter. It represents a natural frame of reference, or a natural coordinate system, within which we can define infall and the assembly of matter. Hence, it allows us to identify preferred directions in the Universe as opposed to other techniques that may be sensitive to scale but are insensitive to directions, such as the two-point correlation function. However, it has been showed that the two-point correlation function can be modified to account for this anisotropy \citep[see][]{2008MNRAS.389.1127P}. Still, the cosmic web provides an essential framework for the study of galaxy and structure formation \citep[see][]{2007MNRAS.381...41H, 2014MNRAS.441.2923C, 2021A&A...651A..56G}. 

The distribution of matter in the Universe can be classified into four components of the cosmic web, which are knots, filaments, sheets and voids. Several algorithms and methodologies have been developed to derive this classification, either based on observational data [e.g. multiscale morphology filter \citep{2007A&A...474..315A, 2014MNRAS.440L..46A}; SpineWeb \citep{2010ApJ...723..364A}; local skeleton \citep{2008ApJ...672L...1S}; FINE \citep{2010MNRAS.407.1449G}; DisPerSE \citep{2011MNRAS.414..350S}; T-Rex \citep{2020A&A...637A..18B}; ORIGAMI \citep{2012ApJ...754..126F, 2015MNRAS.450.3239F}; minimal spanning tree \citep{2014MNRAS.438..177A}; Bisous \citep{2014MNRAS.438.3465T, 2016A&C....16...17T}, multi-stream web analysis \citep{2015MNRAS.452.1643R}, and Lagrangian classifiers \citep{2017JCAP...06..049L}], or numerical data [e.g. Hessian based methods \citep{2007MNRAS.375..489H}; the tidal shear tensor \citep{2009MNRAS.396.1815F}; the velocity shear tensor \citep{2012MNRAS.425.2049H}; CLASSIC \citep{2012MNRAS.425.2443K}, or NEXUS \citep{2013MNRAS.429.1286C}]. A few of these methods have been brought together and compared in \cite{2018MNRAS.473.1195L}.

Considering the notion of cosmic web as a description of the anisotropic mass assembly of matter, \cite{2014MNRAS.443.1274L} showed that the preferential direction of subhalo accretion is aligned with the axis of weakest collapse of the velocity shear tensor. \cite{2015MNRAS.450.2727T} discussed the role of the cosmic web in the accretion of satellites using observational data from the Sloan Digital Sky Survey \citep[SDSS,][]{2011ApJS..193...29A}. \cite{2015ApJ...813....6K} also studied subhalo accretion within the context of the cosmic web, focusing on the alignment with halo shape. Besides, \cite{2018MNRAS.473.1562W} explored the correlation between the halo spin and the large scale structures using the cosmic web, and later on showed that the spins of low-mass galaxies are preferentially aligned with the slowest collapsing direction, i.e the eigenvector $\vec{e_3}$ of the tidal velocity field \citep{2018ApJ...866..138W}. In \cite{2018A&A...613A...4W}, the authors aimed to explore the connection of the anisotropy of the spatial distribution of satellite galaxies to the local cosmic web.
Finally, \cite{2020ApJ...900..129W} considered group and filaments data from SDSS and found that satellites are accreted along filaments \citep[see also][]{2020A&A...635A.195G}.

Many studies have related the infall pattern of satellites onto the Local Group to the flattened distribution of satellites of the Milky Way and Andromeda. \cite{2004ApJ...603....7K} suggested that the origin of the flattened distribution of satellites is linked to a non-uniform infall pattern of accreted subhaloes. By examining the infall of subhaloes using a constrained simulation of the Local Group (LG), hence reproducing the essential properties of the LG, \cite{2011MNRAS.411.1525L} showed that satellites galaxies tend to be accreted from preferred directions, rather than being accreted uniformly in all directions in the sky. \cite{2017MNRAS.472.4099K} examined the difference in subhalo accretion between cold and warm dark matter cosmologies. Finally, \cite{2018MNRAS.476.1796S} studied the incidence of group and filamentary dwarf galaxy accretion into the Milky Way, using hydrodynamical simulations, and found that the accretion of dwarf galaxies takes place preferentially perpendicular to the halo minor axis. 

In this paper, we focus on investigating whether the preferential direction of subhaloes accretion in the LG is related to the cosmic web and the local environment. We ask the questions: are satellites in constrained simulations anisotropically accreted, and if so, are the directions co-incident with the local cosmic web? We will make use of the HESTIA simulations suite \citep{2020MNRAS.498.2968L} and will consider satellites of both MW and M31. The paper is organized as followed. The methodology section, i.e section \ref{sec:methods}, describes the HESTIA simulations, the halo finder and the merger tree algorithms, as well as the cosmic web and the velocity shear and tidal tensors. The results of this paper are presented in section \ref{sec:results}. The paper ends with a short conclusion in section \ref{sec:conclusion}, and an appendix \ref{sec:appendix} showing tests conducted on lower resolution simulations, which are also part of the HESTIA suite.

\section{Methods}
\label{sec:methods}

\subsection{HESTIA: simulations of the Local Group}
\label{sec:sims}

The HESTIA simulation suite \citep{2020MNRAS.498.2968L} is a set of low-resolution and high-resolution cosmological magnetohydrodynamical simulations of the Local Group (LG). 

The \textit{Cosmicflows-2} (CF2) peculiar velocities \citep{2013AJ....146...86T} are used to construct the initial conditions (within the framework of constrained simulations). The CF2 catalog is grouped \citep{2018MNRAS.476.4362S} and bias-minimized \citep{2015MNRAS.450.2644S}. The simulations are built using the \texttt{AREPO} code to solve the ideal magnetohydrodynamics (MHD) equations \citep{2010MNRAS.401..791S,2016MNRAS.455.1134P,2020ApJS..248...32W}, and the AURIGA galaxy formation model \citep{2017MNRAS.467..179G}, which is based on the Illustris model \citep{2013MNRAS.436.3031V} and implements various physical processes that are the most relevant for the formation and evolution of galaxies. \cite{2014MNRAS.444.1518V, 2014Natur.509..177V} uses the same galaxy formation model. The reader can also refer to \cite{2020NatRP...2...42V} for a review on cosmological simulations.

The entire suite of the HESTIA simulations assumes a $\Lambda$CDM cosmology with the following \cite{2014A&A...571A..16P} values: $\sigma_8 = 0.83$, $H_0 = 100h$ km s$^{-1}$ Mpc$^{-1}$ where $h = 0.677$, $\Omega_\Lambda = 0.682$, $\Omega_m = 0.270$ and $\Omega_b = 0.048$.

The publicly available \texttt{AHF} halo finder \citep{2009ApJS..182..608K} identifies haloes, subhaloes, and their properties, at each redshift by detecting gravitationally bound particles. Structures containing less than 20 bound particles are not considered. Halo histories are derived by the \texttt{MERGER TREE} algorithm, a package included in the AHF software. 

In this paper, we consider three Local Groups simulated at high resolution, namely \texttt{09\_18}, \texttt{17\_11}, \texttt{37\_11} (the numbers are based on the random seed used). These three simulations each consists of a region of two 3.7 Mpc spherical volumes centered on the two primary haloes at $z=0$, filled with $8192^3$ effective particles, achieving a mass and spatial resolution of $m_\mathrm{dm} = 1.5 \times 10^5$ M$_\odot$, $m_\mathrm{gas} = 2.2 \times 10^4$ M$_\odot$ and $\epsilon = 220$ pc. A larger number of lower resolutions simulations are considered in the appendix.

The reader may refer to \cite{2020MNRAS.498.2968L} for more details about the HESTIA suite and the algorithms mentioned above.

\subsection{The cosmic web: velocity shear and tidal tensors}

To evaluate the cosmic web, we consider the velocity shear tensor, derived from the distribution of the particles in the simulation box. First, a clouds-in-cell (CIC) technique is applied to the simulation to compute the density and velocity fields in a $256^3$ grid of side length $100$ Mpc/$h$, hence resulting in a spatial resolution of $0.39$ Mpc/$h$. The grid is then smoothed with a Gaussian kernel, and various smoothing lengths $r_s = 1, 2, 5$ Mpc/$h$ are considered. 

Finally, the shear in the velocity field is derived from the velocity field as \citep{2012MNRAS.425.2049H}:
\begin{equation}
    \Sigma_{\alpha\beta} = - \frac{1}{2 H_0} \left( \frac{\partial V_\alpha}{\partial r_\beta} + \frac{\partial V_\beta}{\partial r_\alpha} \right),
    \label{eq:shear}
\end{equation}
where $H_0$ is the Hubble constant, and the partial derivatives of the velocity $V$ are derived along the directions $\alpha$ and $\beta$, representing the orthogonal Supergalactic Cartesian axes $x$, $y$ and $z$. In each grid cell, we obtain the eigenvectors $\vec{e_1}$, $\vec{e_2}$ and $\vec{e_3}$, and the corresponding eigenvalues ordered such that $\lambda_1 > \lambda_2 > \lambda_3$.

We also consider in this work the tidal shear tensor \citep{1970A&A.....5...84Z,2007MNRAS.375..489H}, defined as the Hessian of the gravitational potential $\phi$:
\begin{equation}
    T_{\alpha\beta} = - \frac{\partial^2\phi}{\partial r_\alpha \partial r_\beta},
\end{equation}
where the gravitational potential is rescaled by a factor $4\pi G \bar\rho$ (where G is the gravitational constant and $\bar\rho$ is the mean density of the Universe), and obeys the Poisson equation $\nabla^2 \phi = \delta$, where $\delta$ is the matter overdensity field.

Throughout this paper and for both tensors, the eigenvectors are computed following the methodology described in \cite{2020NewA...8001405W}, which differs from the usual Nearest Grid Point method, as in this case the Hessian matrix is computed and solved for each halo in the simulation, instead of doing it once for the entire grid. In practice the tensor fields are first evaluated on a gird. Instead of then begin evaluated on the grid, the method of Wang et al 2020 computes a single tensor associated to each halo {\it before} diagonalisation. The tensor field associated with each halo is constructed based on the 7 adjoining cells in a CIC inspired way, namely weighted by distance to the halo in consideration. Each halo thus has a unique velocity or tidal tensor associated with it which is then diagonalized and whose eigenvectors are then used.

\subsection{Identifying accreted subhaloes}

AHF uses the parameter $R_{200}$\footnote{The parameter $R_{200}$ is defined as the radius at which the mean enclosed matter density is $\rho (<R_{200}) = 200 \times \rho_\mathrm{crit}$, where $\rho_\mathrm{crit}$ corresponds to the critical density for closure.} to determine masses and radii of objects. In this case, the actual virial radius $R_\mathrm{vir}$ of halos sits between $R_{200}$ and $2 \times R_{200}$. Hence, except when specifically mentioned, we consider the moment of infall at $2 \times R_{200}$, as $R_\mathrm{vir}$ is larger than $R_{200}$ and we do not expect much signal at that threshold (for reasons that will be explained and examined below).

Satellite galaxies are identified as being within $R_{200}$ of their main host at $z=0$. These are then tracked backwards in time by following the main branch of the merger tree, which tracks the location of the most massive progenitor (sub)halo. A moment of accretion is identified as when the satellite first crossed $2R_{200}$. This is termed $z_{inf}$. For clarity we repeat that satellites are identified as those haloes that reside within $R_{200}$ at $z=0$ but their infall time is when they first crossed $2R_{200}$. This is because we are interested in examining the origin of the classical satellite population (not all LG dwarfs) and if their accretion happened at earlier times.

This is done by examining two adjacent snapshots. A subhalo is deemed to have been accreted if in the later snapshot it is located within a sphere of radius
$2 \times R_{200}$ centered on a main progenitor of one of the two LG haloes (MW or M31), where $R_{200}$ is defined by the corresponding main progenitor. The progenitor of the subhalo has to be located outside of the so defined sphere at the earlier snapshot. By going backwards from $z=0$ through all available snapshots, two at a time, we can follow the trajectories of satellites and identify the place and time of accretion. As mentioned earlier, only the subhaloes that survive up to $z=0$ (i.e the ones that are present within $R_{200}$ in the $z=0$ snapshot), are considered. Moreover, we only look at the first time the subhaloes cross the $2 \times R_{200}$ threshold.

We refer to the time of infall as the last snapshot, corresponding to the redshift $z_\mathrm{inf}$, before a given subhalo crosses for the first $2 \times R_{200}$. We can then define the direction of infall $\vec{r}_\mathrm{inf}$ as the coordinates of the subhaloes at infall, re-centered on the respective main host (MW or M31). We also note the total mass of a subhalo at infall $M_{200}^\mathrm{inf}$.

\section{Results}
\label{sec:results}

\subsection{Properties of accretion onto the Local Group}

\begin{figure}
\centering
\includegraphics[width=0.4\textwidth]{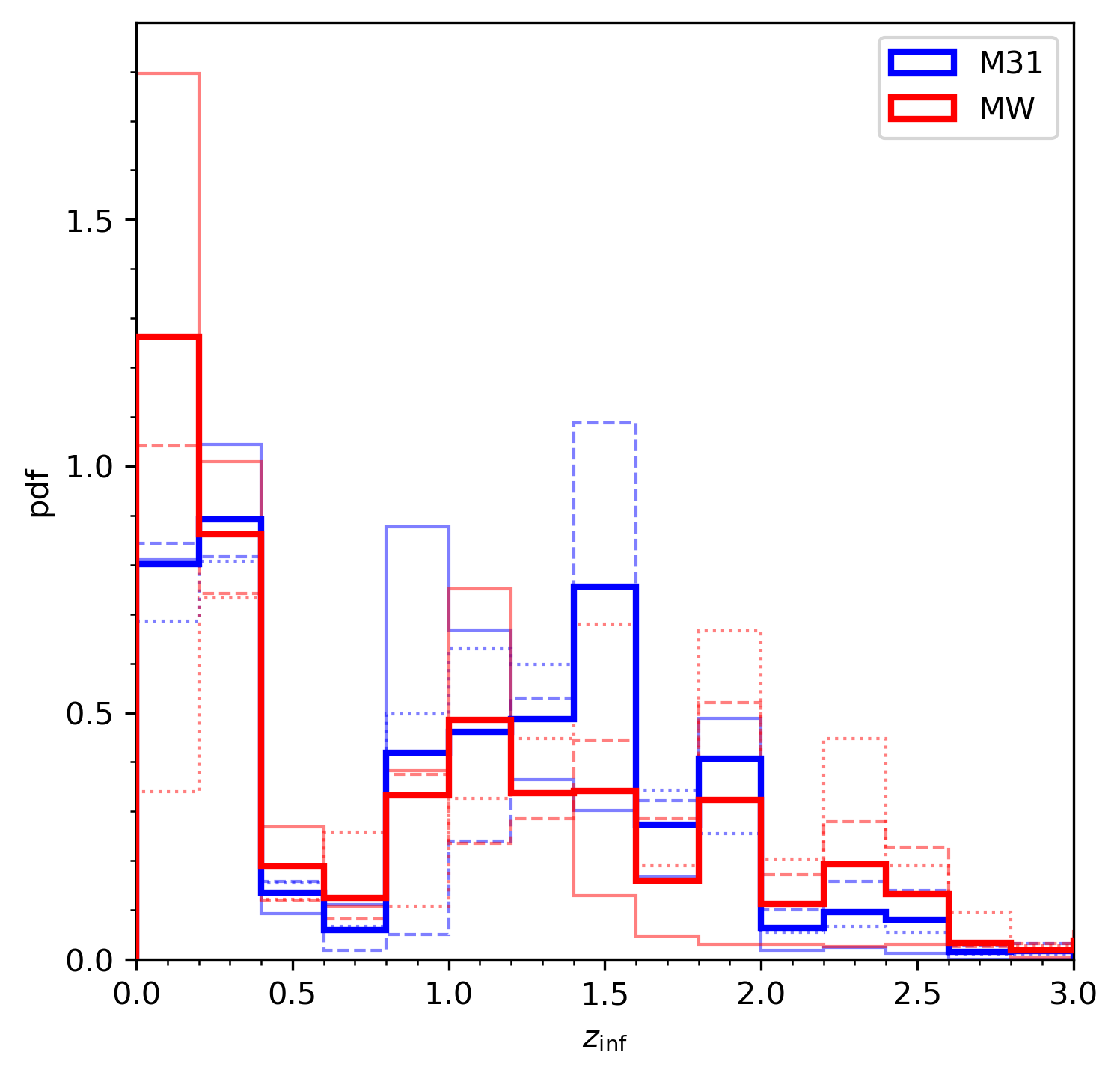}
\caption{Distribution of redshifts at infall $z_\mathrm{inf}$ of all satellites that went through $2 \times R_{200}$ of M31 (blue)  or the MW (red), quantified as a probability distribution function. Thin solid, dashed and dotted lines represent each one realization of the Local Group. The bold solid lines give the total infall redshift distribution, merging all three simulations.} 
\label{fig:zhist}
\end{figure}

The distribution of the redshifts $z_\mathrm{inf}$, i.e when the accreted subhaloes cross $2 \times R_{200}$ of the MW (red lines) or M31 (blue lines), is shown in Figure \ref{fig:zhist}. The three LGs considered are represented by thin solid, dashed or dotted lines (\texttt{09\_18}, \texttt{17\_11} and \texttt{37\_11} respectively). The bold lines show the total redshift distribution of all satellites taken all together, for each main halo. Two peaks can be observed: late infall and early infall. The majority of satellites are accreted recently, after $z_\mathrm{inf} = 0.7$. Other subhaloes get accreted much earlier $z_\mathrm{inf} > 0.7$ and survive up to $z = 0$. Also, we observe that the three individual LGs show a similar trend. This demonstrates the quality and power of the constrained simulations, as such a behavior would not be expected from randomly selected LGs. In fact it has been previously established \citep{2016MNRAS.460.2015S, 2010MNRAS.401.1889L, 2020MNRAS.491.1531C} that one of the main traits of constrained simulations is that they limit the assembly histories of specific objects. We  also note that both peaks of the distributions (total and for each LG) of the two main hosts overlap. This indicates that it is a feature of the entire LG, and that the haloes of the MW and M31 have likely had similar accretion histories.

\begin{figure}
\centering
\includegraphics[width=0.4\textwidth]{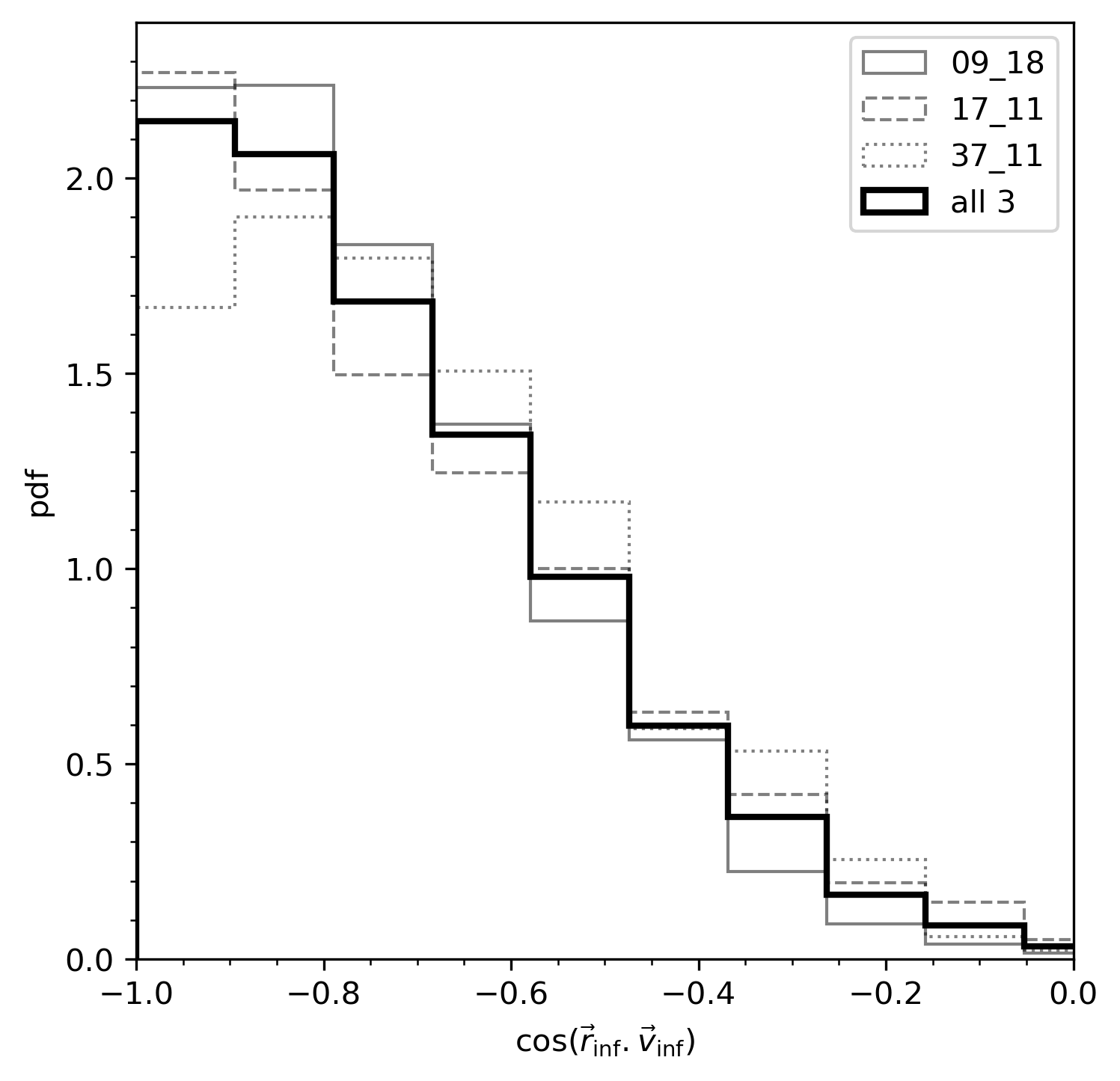}
\caption{Distribution of the cosine of the angle between the direction of infall $\vec{r}_\mathrm{inf}$ and the velocity vector at infall $\vec{v}_\mathrm{inf}$, by means of a probability distribution function. Thin solid, dashed and dotted lines represent each one realization of the Local Group. The bold solid lines gives the total infall redshift distribution, merging all three simulations.}
\label{fig:vdotr}
\end{figure}

Figure \ref{fig:vdotr} shows the distribution of the cosine of the angle between the infall direction $\vec{r}_\mathrm{inf}$, vector at infall of the accreted satellite galaxy and the velocity vector at infall $\vec{v}_\mathrm{inf}$. Both of these vectors are centered on the parent halo's frame. Thin solid, dashed and dotted lines correspond to a single realization of the Local Group. The bold solid lines gives the total distribution, combining all three simulations together. We can observe that the distribution peaks around $\cos{(\vec{r}_\mathrm{inf} . \vec{v}_\mathrm{inf})} = -1$, i.e, the two vectors are aligned with opposite direction, meaning that the satellites are accreted radially, along their velocity.

\begin{figure}
\centering
\includegraphics[width=0.4\textwidth]{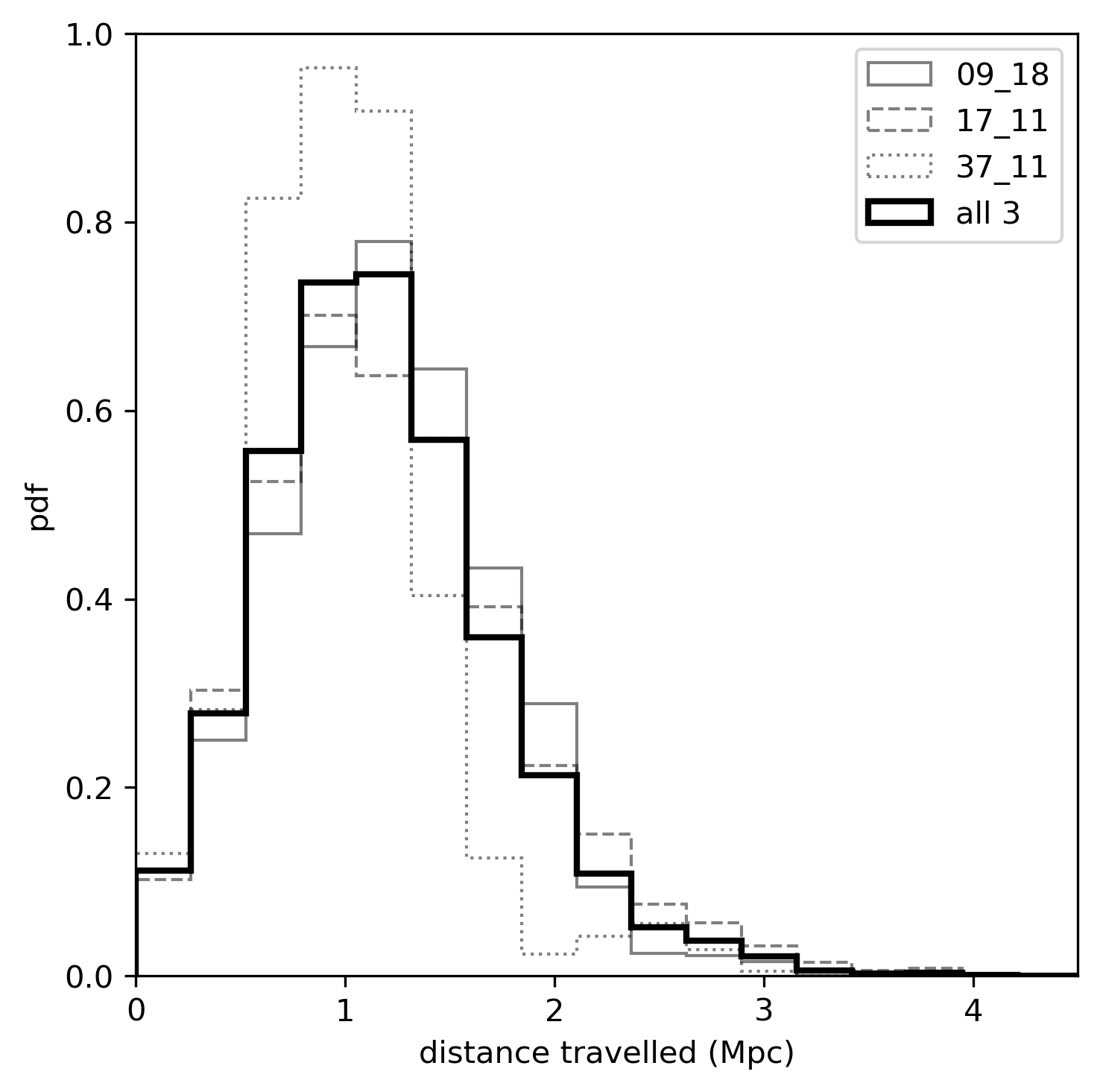}
\caption{Distribution of the minimum distance travelled within the rest frame of the host, i.e the distance in co-moving coordinates between the position at birth (at first recorded snapshot) and the position at accretion, re-centered on the position of the main halo. The distribution is represented by a probability distribution function.}
\label{fig:dhist}
\end{figure}

We can approximate the pre-infall distance travelled by satellites by considering the co-moving coordinate distance between their position at the first snapshot recorded (i.e, at {\it birth}) and their position at $z_\mathrm{inf}$, both re-centered on the respective host halo. This approximation gives the minumum distance travelled by satellites before accretion, as we assume that their path from birth to infall is a straight line, and do not consider their actual trajectories. The distribution of the  distance travelled within the host rest frame, combining all 3 simulations, is shown in Figure \ref{fig:dhist}. We observe that subhaloes have travelled a distance up to 4 Mpc before getting accreted. More importantly, we notice a peak at 1 Mpc, showing that on average most $z=0$ subhaloes traveled a distance of roughly 1 Mpc before crossing the $2 \times R_{200}$ threshold of the MW or M31. That said, the median of the distribution is greater than this, implying that the LG accretes from, roughly, a sphere of up to 4 Mpc around the two main hosts. 

\begin{figure}
\centering
\includegraphics[width=0.4\textwidth]{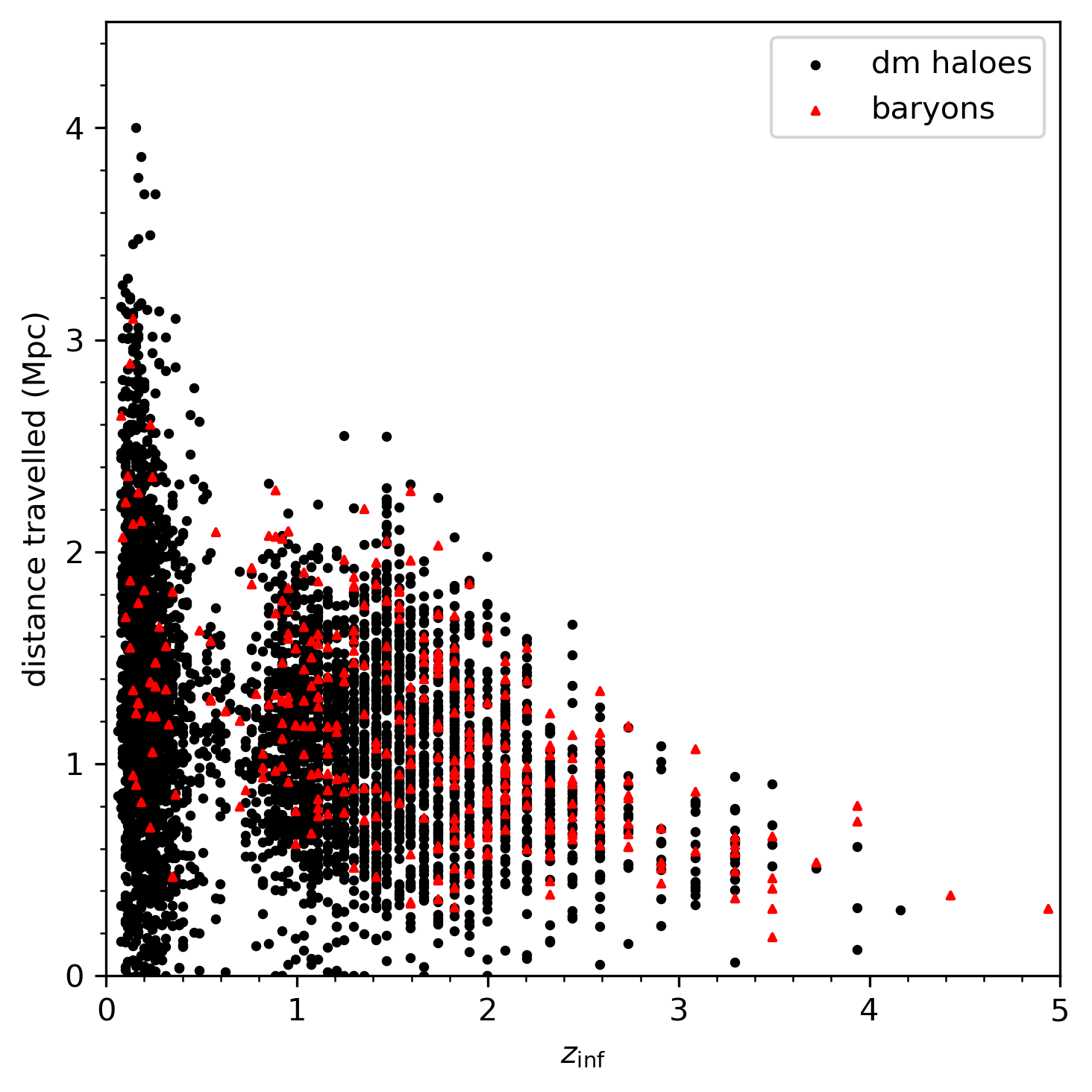}
\caption{Minimum distance travelled as function of the redshift at infall $z_\mathrm{inf}$. The blue and red square represent the dark matter and baryon accreted subhaloes, respectively.}
\label{fig:dzinf}
\end{figure}

In Figure \ref{fig:dzinf}, the minimum co-moving distance travelled described above is plotted as a function of the redshift at time of infall $z_\mathrm{inf}$. Dark matter only haloes are represented by black dots, while haloes containing baryons are highlighted by red triangles. Firstly, one can easily notice the two peaks in the redshift distribution as seen in Figure \ref{fig:zhist}, before and after $z=0.7$. Moreover, the decreasing trend shows that the satellites accreted recently (recent $z_\mathrm{inf}$) are the one that travelled the most, i.e with the largest minimum travelled distance. This is because the haloes that were recently accreted have had the most amount of time to travel across the Universe. Those accreted at early times had less time at their disposal and thus were not able to cross large distances. 

\subsection{Anisotropic infall within the context of the cosmic web}

In this section, we analyze the alignment of the infall direction $\vec{r}_\mathrm{inf}$ with the three eigenvectors of the velocity shear and tidal tensors, by considering the distribution of the cosine of the angle between $\vec{r}_\mathrm{inf}$ and $\vec{e_1}$, $\vec{e_2}$ and $\vec{e_3}$. As eigenvectors are non-directional, $0 \leq \cos{(\vec{r}_\mathrm{inf}.\vec{e}_{1,2,3})} \leq 1$. A cosine of 1 means that the two directions are parallel while 0 implies they are perpendicular. To quantify the strength of the distribution of $\cos{(\vec{r}_\mathrm{inf}.\vec{e}_{1,2,3})}$, we calculate its statistical significance as the difference between the actual distribution and the median of 10,000 uniform distributions of the same size, averaged over all bins and calculated in units of the Poisson error. The uniform distributions are derived from 10,000 random samples of points uniformly distributed on a sphere. High values of the statistical significance indicate a strong deviation from a uniform distribution, hence a strong signal, while low values show a statistically weak signal, close to or consistent with a uniform distribution. 

\begin{figure*}
\centering
\includegraphics[width=0.75\textwidth]{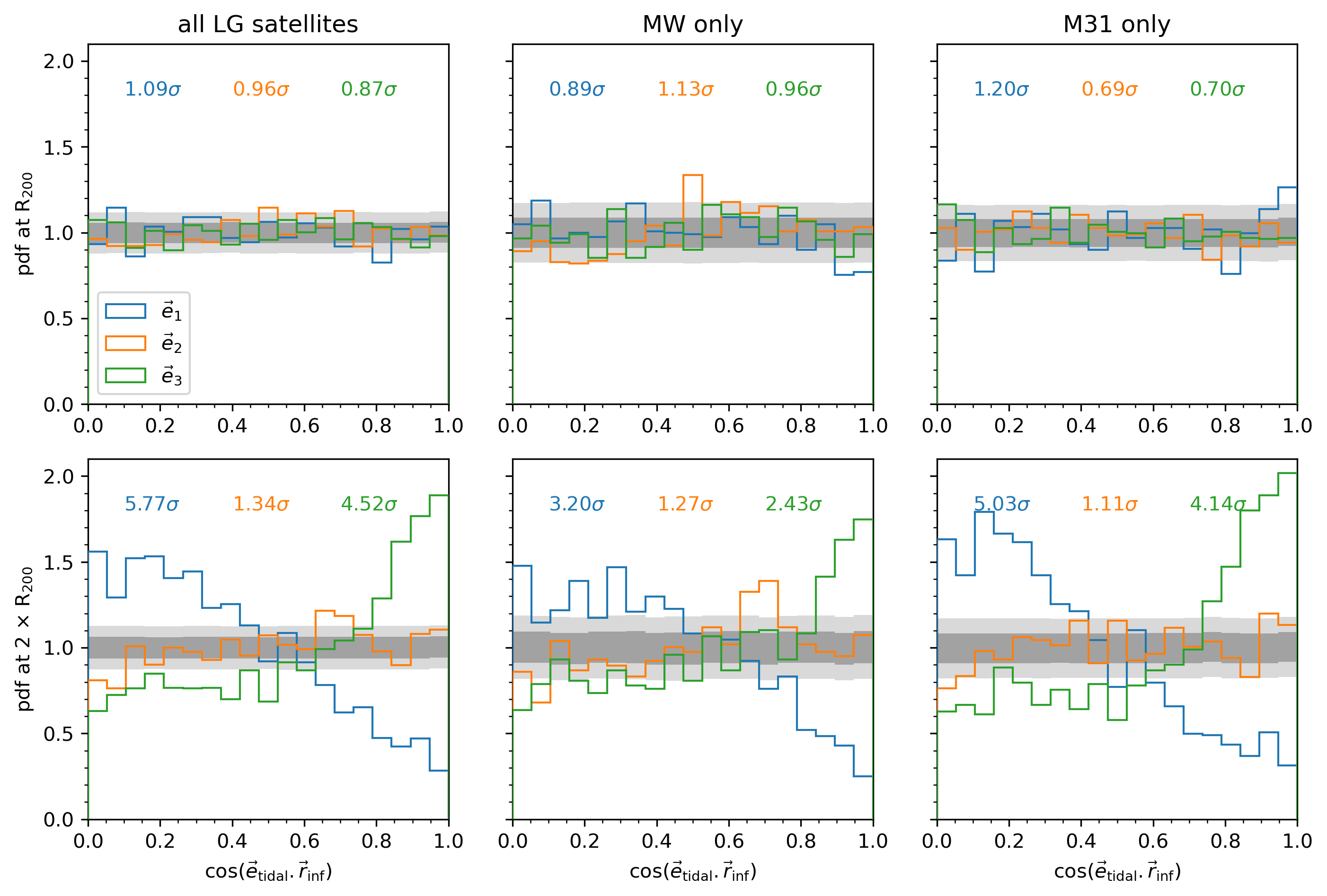}
\caption{Distribution of the angles between the infall direction $\vec{r}_\mathrm{inf}$ at $R_{200}$ (top row) and $2 \times R_{200}$ (bottom row) and the three eigenvectors $\vec{e_1}$ (blue), $\vec{e_2}$ (orange) and $\vec{e_3}$ (green) of the tidal tensor, where a smoothing of 1 Mpc has been applied. All distributions are depicted by probability density distributions. The three columns depict the distribution of all accreted satellites (all three high resolution simulations combined), MW satellites only and M31 satellites only, from left to right respectively. The statistical significance of each distribution is depicted on the top of each panel, and is characterized by the average offset in each bin between the considered distribution and the median of 10,000 uniform ones, calculated in units of the Poisson error. The shaded light and dark grey region represent respectively the $\pm 1 \sigma$ and $\pm 2 \sigma$ thresholds, where $\sigma$ is the standard deviation of 10,000 uniform distribution. The reader may check the corresponding text in the manuscript for more details.}
\label{fig:infalldir}
\end{figure*}

Figure \ref{fig:infalldir} shows the distribution of $\cos{(\vec{r}_\mathrm{inf}.\vec{e}_{1,2,3})}$ at $R_{200}$ (top row) and $2 \times R_{200}$ (bottom row). Note this is the only plot in the paper where we have examined the accretion at $R_{200}$. The three eigenvectors $\vec{e_1}$, $\vec{e_2}$ and $\vec{e_3}$ are represented respectively by blue, orange and green lines. Both tidal and shear tensors have been checked, however, the obtained distributions being similar, only the tidal tensor eigenvectors are shown for clarity. Similarly, as all smoothing lengths applied to the tensors show similar trends, only the distributions obtained with a smoothing length of $r_s=1$ Mpc are displayed. From left to right columns depict all satellites (combining all three simulations and both host haloes), MW satellites only, and M31 satellites only, respectively. The statistical significance of each distribution is noted on the top of each panel, with the same color code as their respective eigenvector: blue for $\vec{e_1}$, orange for $\vec{e_2}$, and green for $\vec{e_3}$. The grey shaded regions depict the Poisson error, i.e the standard deviation $\sigma$ of 10,000 uniform distributions at each bin. The dark grey shows the $\pm 1 \sigma$ interval, while the lighter grey shows the $\pm 2 \sigma$ interval. 

The distributions in top row panels of Figure \ref{fig:infalldir} being mostly located within the grey regions, and the values of the statistical significance being mostly less or close to unity, indicate there is no significant alignment of the infall direction with any eigenvector of the tidal tensor. We remind the reader that the top row corresponds to the distribution of accreted subhaloes at $R_{200}$. However, when looking at the distributions at $2 \times R_{200}$ in the bottom row panels of Figure \ref{fig:infalldir}, we notice that the infall direction $\vec{r}_\mathrm{inf}$ is strongly aligned with the axis of slowest collapse $\vec{e_3}$. We also notice that $\vec{r}_\mathrm{inf}$ is orthogonal to $\vec{e_1}$, while no significant alignment is observed with $\vec{e_2}$. At $R_{200}$, i.e in the non-linear regime, the signal is extremely weak due to non-linear dynamics. Beyond $R_{200}$, we transition from the virial regime to the quasi-linear (QL) regime, which explains why the signal is much stronger at $2 \times R_{200}$.

Besides, when comparing the middle (MW satellites) and right (M31 satellites) columns of Figure \ref{fig:infalldir}, corresponding to satellites accreted by each host separately, no significant difference can be observed in the top row, i.e at $R_{200}$, for the same reasons described above. At $2 \times R_{200}$, the distributions corresponding to both hosts look similar, however the statistical significance corresponding to the satellites accreted by M31 is $\sim 1.7$ times higher than the one corresponding to the MW satellites. This shows that, although the alignment with $\vec{e_3}$ is observed for both hosts, it is stronger for M31, and weaker for the MW satellites. Such a difference may be explained by the fact that M31 is more massive than the MW \footnote{The more massive halos are called M31 by definition in HESTIA simulations.}.

In the next figures, we look at the distribution of $\cos{(\vec{r}_\mathrm{inf}.\vec{e}_{1,2,3})}$ at $2 \times R_{200}$ split by mass and redshift, in order to check for any dependence. 

\begin{figure*}
\centering
\includegraphics[width=0.75\textwidth]{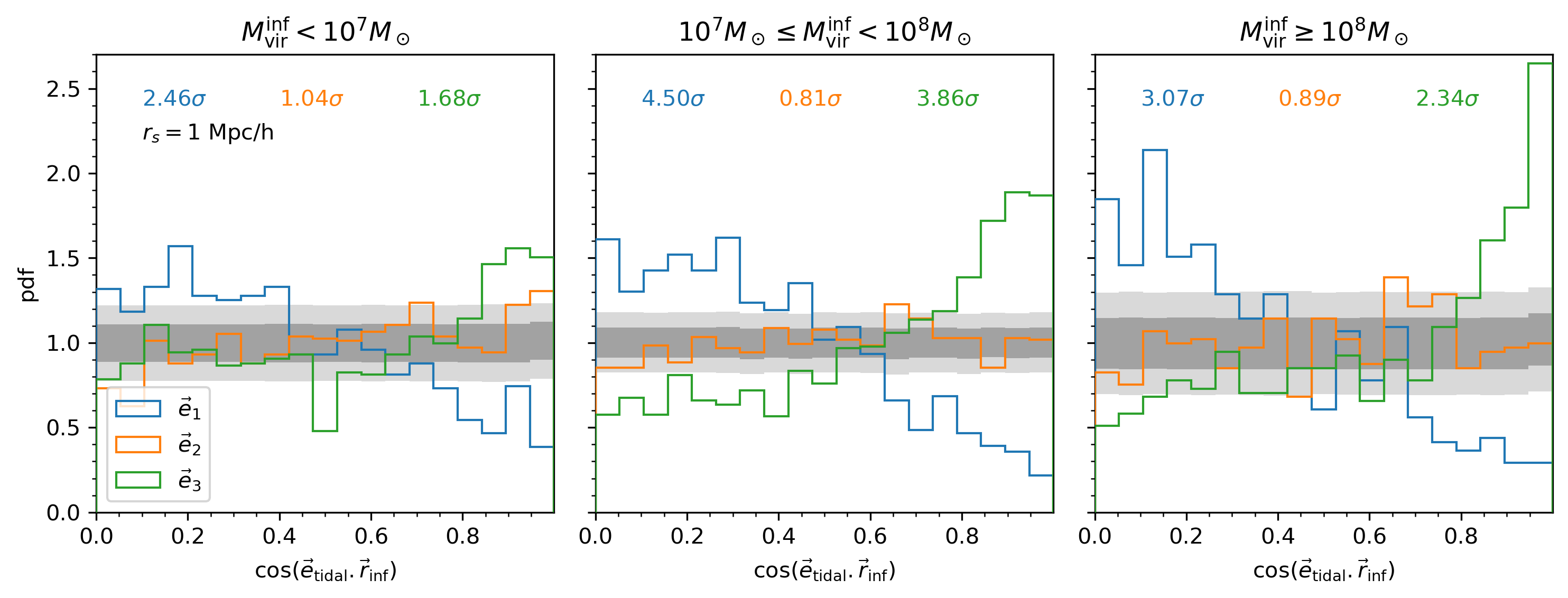}
\caption{Distribution of $\cos{(\vec{r}_\mathrm{inf}.\vec{e}_{1,2,3})}$ at $2 \times R_{200}$, split by total mass of satellites at infall $M_{200}^\mathrm{inf}$ (first column: $M_{200}^\mathrm{inf} < 10^7$ M$_\odot$, second column: $10^7 \leq M_{200}^\mathrm{inf} < 10^8$ M$_\odot$, third column: $M_{200}^\mathrm{inf} > 10^8$ M$_\odot$). The reader may refer to the caption of Figure \ref{fig:infalldir} for the description of the layout and the color code of the different panels.}
\label{fig:mass}
\end{figure*}

First, Figure \ref{fig:mass} shows the distribution of $\cos{(\vec{r}_\mathrm{inf}.\vec{e}_{1,2,3})}$ at $2 \times R_{200}$, where $e_{1,2,3}$ are represented by blue, orange and green lines, respectively. For clarity purposes, only the results obtained from the tidal tensor, smoothed by 1 Mpc, are shown, however both shear and tidal tensors show similar trends. Similarly, the results obtained with other smoothing lengths are not displayed as the distributions are very similar. The distributions are split by total mass of satellites at accretion, namely $M_{200}^\mathrm{inf}$, using the following bins: $M_{200}^\mathrm{inf} < 10^7$ M$_\odot$ (first column), $10^7 \leq M_{200}^\mathrm{inf} < 10^8$ M$_\odot$ (second column) and $M_{200}^\mathrm{inf} > 10^8$ M$_\odot$ (third column). The statistical significance and the standard deviation of a uniform distribution are shown on each panel. The reader may refer to the previous figures for a detailed description. All three mass bins show alignment of the accretion however we observe a mild mass dependence wherein the smallest infall masses ($M_{200}^\mathrm{inf} > 10^7M_{\odot}$) exhibit a weaker signal than the more massive haloes. The  $10^7 \leq M_{200}^\mathrm{inf} < 10^8$ mass bin is stronger.

\begin{figure*}
\centering
\includegraphics[width=0.5\textwidth]{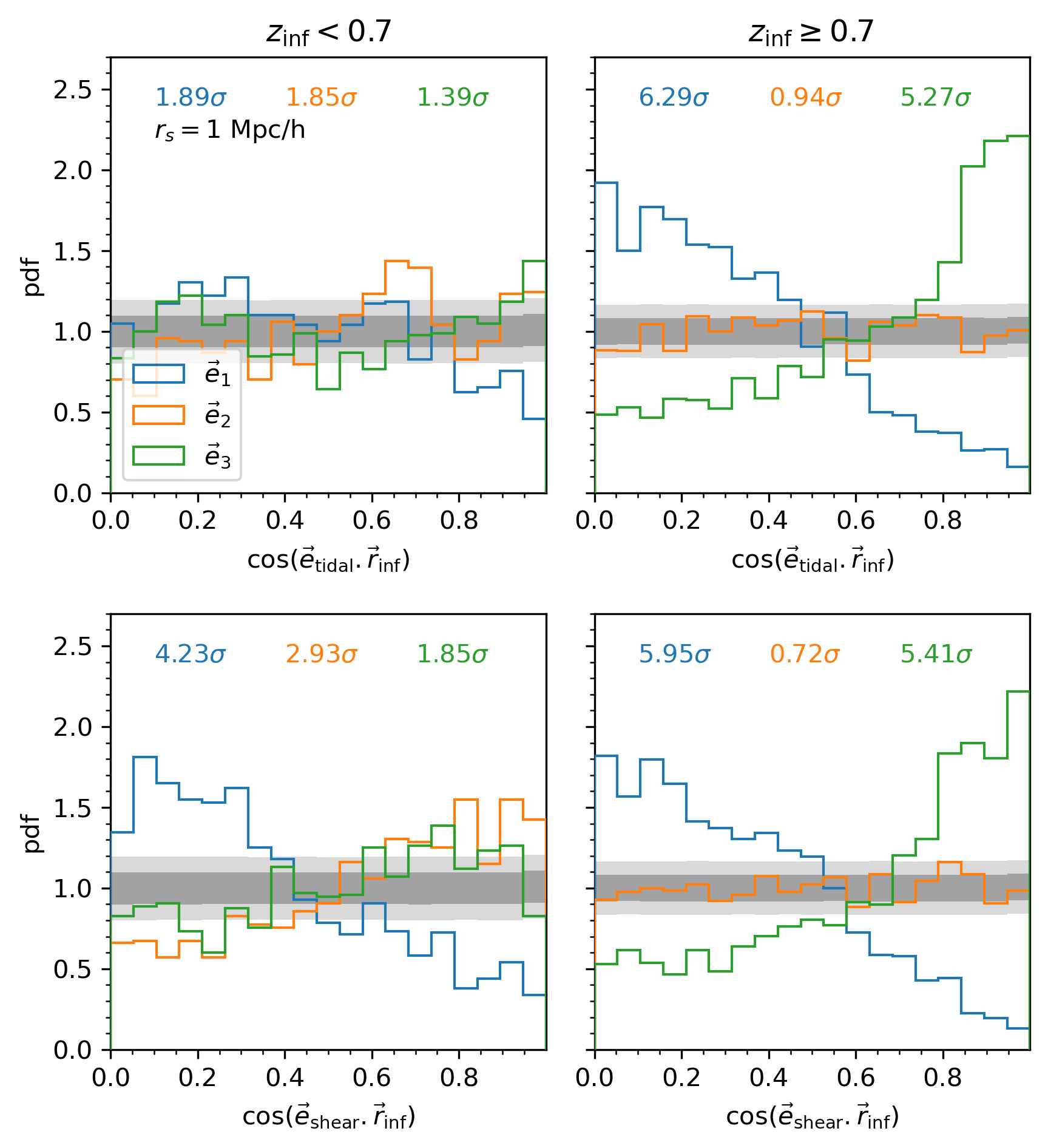}
\caption{Distribution of $\cos{(\vec{r}_\mathrm{inf}.\vec{e}_{1,2,3})}$ at $2 \times R_{200}$, split by redshift at infall (first column: $z_\mathrm{inf} < 0.7$, second column: $z_\mathrm{inf} \geq 0.7$). The upper row corresponds to the tidal tensor, while the bottom row corresponds to the shear tensor. The reader may refer to the caption of Figure \ref{fig:infalldir} for the description of the layout and the color code of the different panels.}
\label{fig:redshift}
\end{figure*}

In Figure \ref{fig:redshift}, the distributions of $\cos{(\vec{r}_\mathrm{inf}.\vec{e}_{1,2,3})}$ at $2 \times R_{200}$ are split into two bins of redshift at infall $z_\mathrm{inf}$: late infall and early infall. Subhaloes accreted recently at $z_\mathrm{inf} < 0.7$ are shown in the left column while the ones accreted  earlier at $z_\mathrm{inf} \geq 0.7$ are shown in the right column. The top row shows the distribution with the tidal tensor's eigenframe while the bottom rows show this distribution for the shear tensor. The reader may refer to the texts of the previous figures for a detailed description of the color code and quantities displayed on the panels. We notice that the alignment with $\vec{e_3}$ is dominated by early infall and is more isotropic at later times.

\begin{figure*}
\centering
\includegraphics[width=0.45\textwidth]{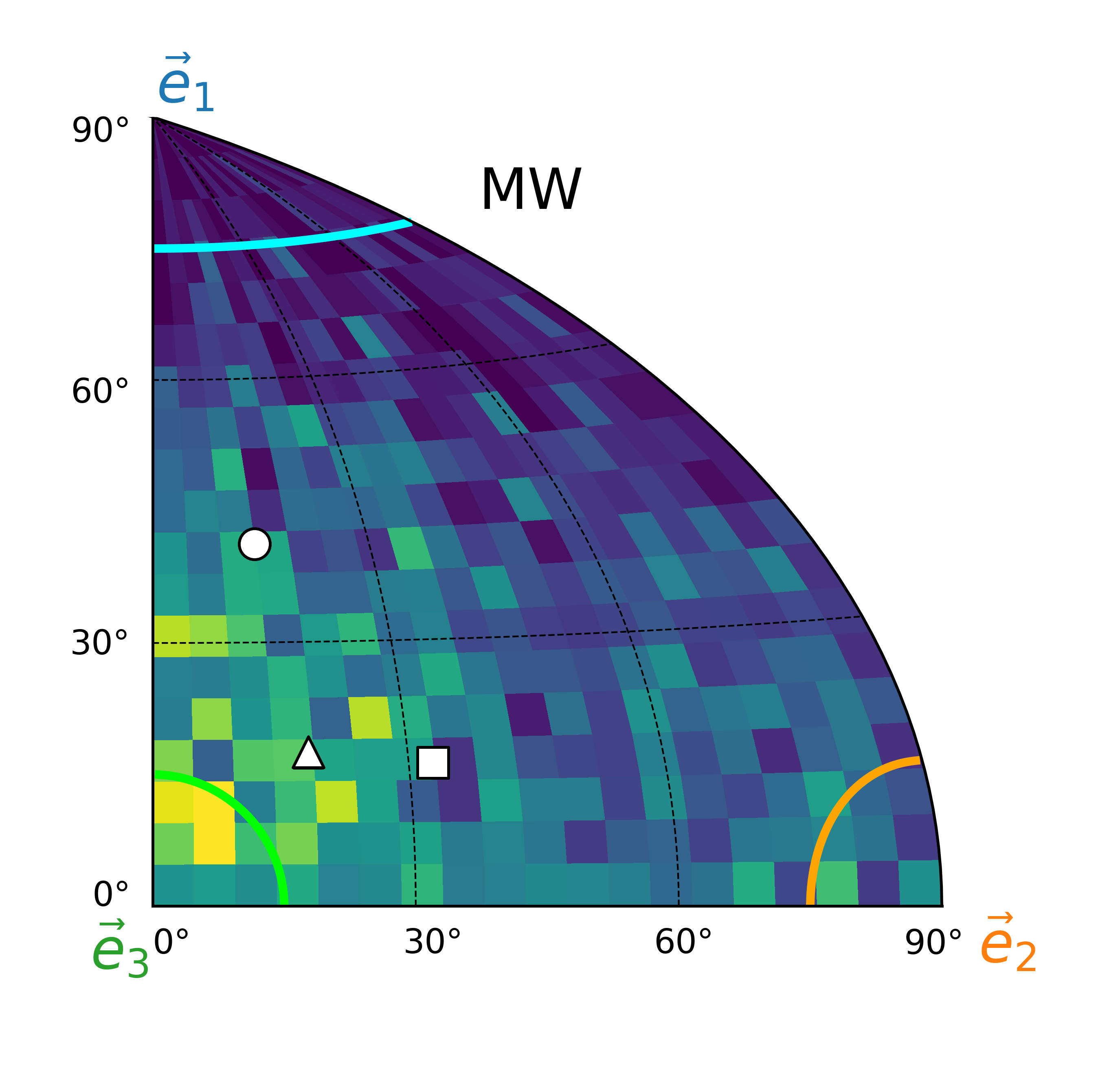}\includegraphics[width=0.45\textwidth]{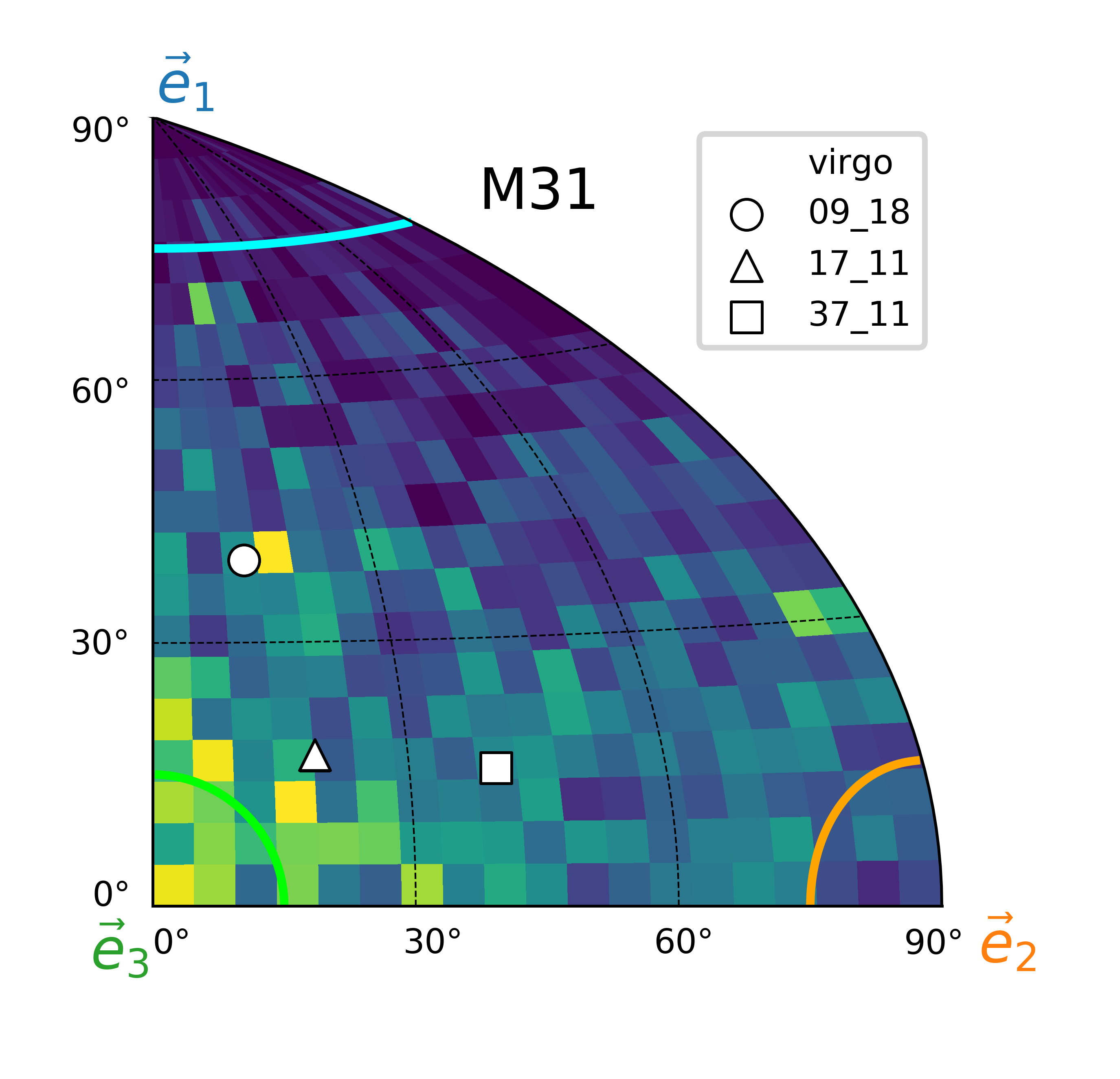}
\caption{Location of MW (left) and M31 (right) satellites entry points as they cross the $2 \times R_{200}$ threshold of their respective main host, shown as the density of points on an Aitoff projection. The projection is oriented in the frame of the tidal tensor eigenvectors smoothed at 7 Mpc, such that the north pole corresponds to the direction of $\vec{e_1}$, the right point on the horizontal axis at $+90\deg$ corresponds to $\vec{e_2}$ and the midpoint corresponds to $\vec{e_3}$. The blue, orange and green circles define areas within $15\deg$ of the eigenvectors $\vec{e_1}$, $\vec{e_2}$ and $\vec{e_3}$, respectively. As the eigenvectors are non-directional, only a single octant of the sphere is shown. Yellow corresponds to a high density of satellite entry points, while dark blue corresponds to a low density of entry points. The location of Virgo simulated in all three realizations is shown as white scattered points (circle: \texttt{09\_18}, triangle: \texttt{17\_11}, square: \texttt{37\_11}).}
\label{fig:octant}
\end{figure*}

Finally, Figure \ref{fig:octant} depicts the location of the MW (left) and M31 (right) satellites entry points as they cross the $2 \times R_{200}$ of their respective main halo. The entry points are shown as a ``heat map'' and plotted on an Aitoff projection of the $2R_{200}$ sphere, oriented in the eigenframe of the tidal tensor, smoothed at 1 Mpc. As the eigenvectors are non-directional, only a single octant of the $2R_{200}$ sphere is shown, instead of an entire full sky projection. The north pole of the projection corresponds to the direction of $\vec{e_1}$, the right point on the horizontal axis at $+90\deg$ corresponds to the direction of $\vec{e_2}$, and the midpoint corresponds to the direction of $\vec{e_3}$. Areas within $15\deg$ of the eigenvectors $\vec{e_1}$, $\vec{e_2}$ and $\vec{e_3}$ are defined respectively by blue, orange and green circles. The yellow pixels correspond to a high density of satellite entry points, while dark blue pixels correspond to a low density of entry points. The location of the simulated Virgo in all three realizations is shown as white scattered points (circle: \texttt{09\_18}, triangle: \texttt{17\_11}, square: \texttt{37\_11}). In both cases, we observe a high density of points around $\vec{e_3}$, which is in accordance with the results shown above. Besides, the high density around the axis of slowest collapse looks more spherical in the case of M31 than for the MW, confirming the different trends between the two main hosts observed in the middle and right panels of Figure \ref{fig:infalldir}. 

\begin{figure*}
\centering
\includegraphics[width=0.9\textwidth]{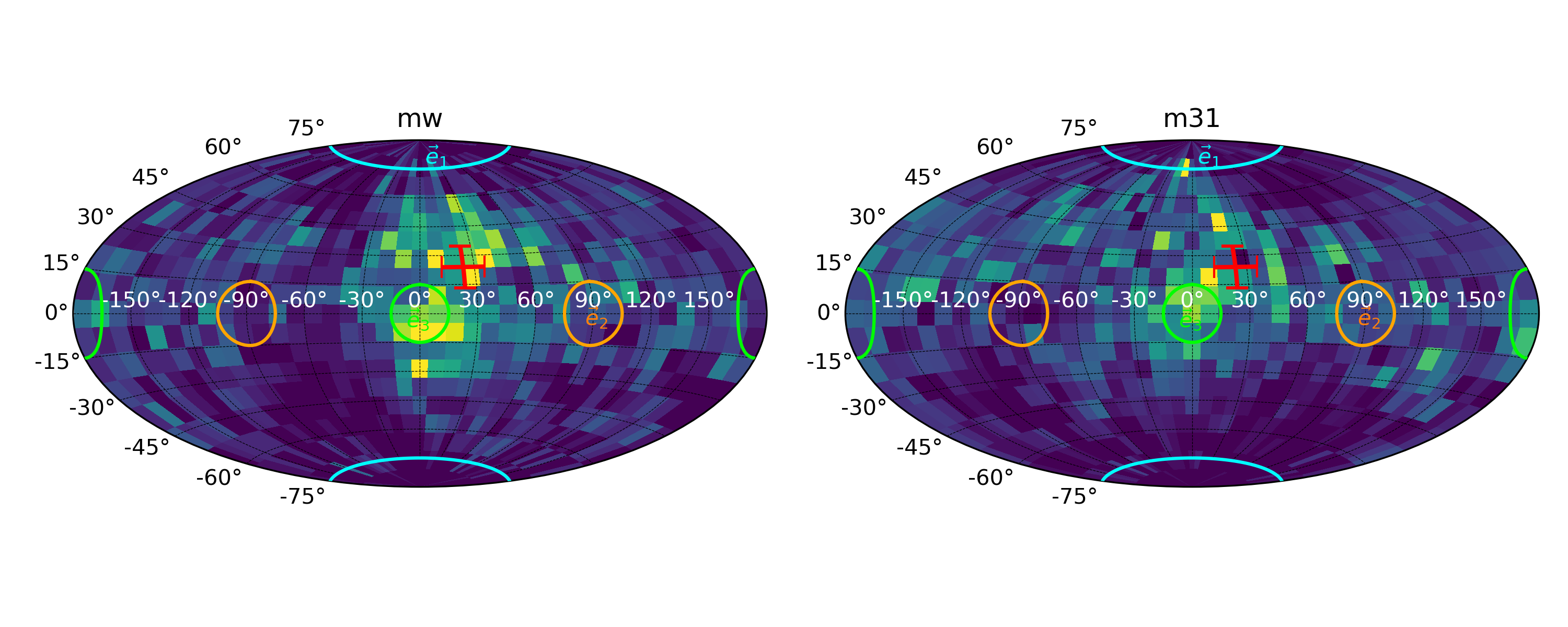}
\caption{Figure \ref{fig:octant} shown as a full sky projection by giving directions to the eigenframe with respect to Virgo. The colormap corresponds to high (yellow) and low (blue) densities of MW (left) and M31 (right) subhalo entry points. Areas within $15\deg$ of the eigenvector are highlighted by blue ($\vec{e_1}$), orange ($\vec{e_2}$) and green ($\vec{e_3}$) circles. The red point and error bars correspond respectively to the mean and standard deviation of the locations of Virgo in the three simulations.}
\label{fig:fullsky}
\end{figure*}

We remind the reader that Figure \ref{fig:octant} represents a single octant of the $2R_{200}$ sphere as eigenvectors are non-directional. However one can give directions to the eigenvectors in order to get a full sky representation of this sphere by choosing  positive direction. We establish an oriented eigenframe with respect to Virgo, by choosing as positive, the direction of each eigenvector $\vec{e_1}$, $\vec{e_2}$ and $\vec{e_3}$ as the direction closest to the orientation of Virgo (effectively putting Virgo into the 1st octant of the coordinate system). This new setting of the eigenframe is shown in Figure \ref{fig:fullsky}. Similarly to the previous Figure \ref{fig:octant}, Figure \ref{fig:fullsky} shows the location of MW (left) and M31 (right) subhaloes entry points as they cross the $2 \times R_{200}$ threshold of their respective main host. The colormap and colored circles are the same as in \ref{fig:octant}. The red point and error bars correspond respectively to the mean and standard deviation of the locations of Virgo in the three simulations. As in Figure \ref{fig:octant}, the infall pattern indicates that subhaloes appear to be accreted from the general direction of Virgo. Put another way, the $\vec{e}_3$ vector points towards Virgo. 

\section{Conclusion}
\label{sec:conclusion}

This paper presents a study of the accretion of $z=0$ satellites in the Local Group, in the context of the cosmic web. The analysis uses the high-resolution simulations available in the HESTIA simulation suite, constrained with the Cosmicflows-2 cosmography, and also considers the velocity shear and tidal tensors derived from the simulations, as they describe the anisotropic assembly of matter in the Universe. The main results of our study are the following:

\begin{enumerate}[(i)]
    \item We observe two eras in the local group history, corresponding to an early and a late infall, i.e before and after $z \approx 0.7$. This gap is observed in all of the LG simulations we considered in this paper (both lower and higher resolutions). 
    \item Most accreted satellites travel $\sim 1$ Mpc before crossing the $R_{200}$ threshold of their respective main host. Some travel up to 4Mpc.
    \item No significant alignment with the large scale structure is observed at $R_{200}$, due to non-linear dynamics.
    \item The infall direction at $2 \times R_{200}$ is strongly aligned with the eigenvector $\vec{e_3}$ of both velocity shear and tidal tensors. $2 \times R_{200}$ probes the transition from the non-linear to quasi-linear regimes. 
    \item The alignment with $\vec{e_3}$ is stronger for the M31 satellites than the MW satellites, as the statistical significance of the alignment is 1.7 times higher. This could be because the M31 halo is more massive and thus exerts more of an influence on the tidal field.
    \item The alignment with $\vec{e_3}$ is slightly stronger for large infall masses, especially within mass bin $10^7 \leq M_{200}^\mathrm{inf} < 10^8$.
    \item However, it has been shown that the alignment with the axis of slowest collapse is dominated by early infall ($z>0.7$).
\end{enumerate}

As mentioned previously in the paper, this analysis considers only the accreted subhaloes that survive up to $z=0$ and are located within $R_{200}$ of their host. Hence, the satellites that are not present in the last $z=0$ snapshot (they may have been stripped and fell below the simulations resolution limit, or simply merged, but still cross the virial radius at some earlier time) are not considered in this study. As a perspective, one could go further with the analysis by considering all accreted satellites in the LG history. This may allow to confirm the alignment of the preferred infall direction with the cosmic web, and look for other properties, while improving the statistics by increasing the number of considered satellites.

This paper has many cosmological implications, as we found evidence that the local cosmography has influence on the accretion of matter within the Local Group. The mass assembly in the Local Group is dictated by the cosmic web, more precisely the velocity shear and tidal tensors, within a range of influence of up to 4 Mpc. This means that what occurs to the Local Group is determined by what takes places within this range of 4 Mpc. 

The preferred direction of accretion being aligned with the axis of slowest collapse $\vec{e_3}$ shows that the mass being fed into the Local Group may come from a local filament, which itself should be aligned with $\vec{e_3}$ as well \citep{2014MNRAS.437L..11T}. Besides, additionally to showing how the accreted subhaloes are arranged with respect to the eigenframe and the simulated Virgo cluster, the aitoff projections in Figures \ref{fig:octant} and \ref{fig:fullsky} also indicate where Virgo is located with respect to the eigenframe. In particular, we notice that the direction of Virgo is close to the direction of $\vec{e_3}$. 
We may compare the results presented in this paper with \cite{2013MNRAS.436.2096S} and \cite{2017ApJ...850..207S}. Those studies concern the formation of the local environment using numerical action orbit reconstructions, which are based on the same Cosmicflows data constraints as the HESTIA high-resolution simulations considered in this paper. Specifically, \cite{2013MNRAS.436.2096S} focuses on the formation details of the Local Group. We know from their stellar populations that most dwarf galaxies formed very early. Additionally, today a large part of these are located within planar structures. In the large volume that the numerical action method models, \cite{2013MNRAS.436.2096S} found that the motion of the Local Group (shared by MW and M31) is predominantly toward -SGZ and +SGY (in Supergalactic cartesian coordinates), i.e aligned with the directions of the $\vec{e}_1$ and $\vec{e}_3$ eigenvectors, respectively. The authors also found that most of the dwarf galaxies are being drawn into the M31 halo oriented in the plane of the $\vec{e}_2$ and $\vec{e}_3$, favoring $\vec{e}_3$, which is consistent with the main result of our paper. \cite{2013MNRAS.436.2096S} also states that the LG satellites have a common origin, as most of them evacuate from the Local Void. In our case, we are not looking specifically at the Local Void, but we do observe that satellites come from a specific direction. \cite{2017ApJ...850..207S} concentrates on the large scale aspects of the Local Group formation, particularly illustrating the importance of the expansion of the Local Void in the development of the Local Sheet (i.e, establishing the compressive eigenvector direction e1) and the attractive influence of the Virgo Cluster (i.e, establishing the expansive eigenvector direction  $\vec{e}_3$). The infall pattern towards M31 and MW is not aligned with the Local Void, but aligned with the Local Sheet. So, first satellites evacuate from the Local Void to the Local Sheet, then from the Local Sheet to the LG. This is consistent both with our intuition that material collapses along  $\vec{e}_1$ first and  $\vec{e}_3$ last, as well as previous work which shows this \citep{2015ApJ...813....6K}. Tracking back the material which forms satellites at this level, is out of scope for this paper, as we only focus on the direction of accretion of MW and M31 within the context of the cosmic web, and not looking at throughout the orbital history of the accreted satellites, from the Local Void to the Local Sheet and the LG. 

Finally, the study presented in this paper might offer a framework regarding the peculiar geometric spatial distribution of satellites in the Local Group. The identification of a preferred direction, with respect to the cosmic web and the cosmography, and a range of influence of the Local Group, could provide support in addressing this issue.


\section*{Acknowledgements}
This work has been done within the framework of the Constrained Local UniversE Simulations (CLUES) simulations.
AD is supported by a KIAS Individual Grant (PG087201) at Korea Institute for Advanced Study.
YH has been partially supported by the Israel Science Foundation grant ISF 1358/18.
ET acknowledges support by ETAg grant PRG1006 and by the EU through the ERDF CoE grant TK133.
AK is supported by the Ministerio de Ciencia, Innovaci\'{o}n y Universidades (MICIU/FEDER) under research grant PGC2018-094975-C21 and further thanks Claudine Longet for nothing to lose.
HC is grateful to the Institut Universitaire de France and CNES for its support.
JS acknowledges support from the French Agence Nationale de la Recherche for the LOCALIZATION project under grant agreements ANR-21-CE31-0019.
The authors gratefully acknowledge the Gauss Centre for Supercomputing e.V. (www. gauss-centre.eu) for funding this project by providing computing time on the GCS Supercomputer SuperMUC at Leibniz Supercomputing Centre (www.lrz.de).

\section*{Data Availability}

The data used in this work were extracted from the Hestia simulation suite. Requests for access to the Hestia simulation data should be directed to a CLUES Collaboration and will be made available upon reasonable request.


\bibliographystyle{mnras}
\bibliography{A.biblio} 


\appendix

\section{Tests}
\label{sec:appendix}

In order to gauge convergence of the results shown here tests have been carried out on lower resolution simulations. Throughout this section, we consider 13 lower resolution Local Group simulations, filled with 4096$^3$ effective particles, instead of 8192$^3$ as in the main body of the paper. Due to the lower resolution, we expect to identify fewer accreted satellites. However with more simulations we hope to make up for the lack of resolution.  The other parameters stay unchanged, so the reader may refer to section \ref{sec:sims} for a detailed reminder of the simulations parameters. The lower resolution simulations have dark matter particles mass of $1.2 \times 10^{6} M_\odot$, gas mass of $1.8\times 10^{5}M_{\odot}$ and a softening of 340 pc. This means that fewer satellites per host halo are resolved. A total of 1883 satellites (774 MW's and 1109 M31's) are identified in the $4096$ simulation, against 4486 (2122 MW's and 2364 M31's) satellites in the higher resolution one. 

\begin{figure}
\centering
\includegraphics[width=0.4\textwidth]{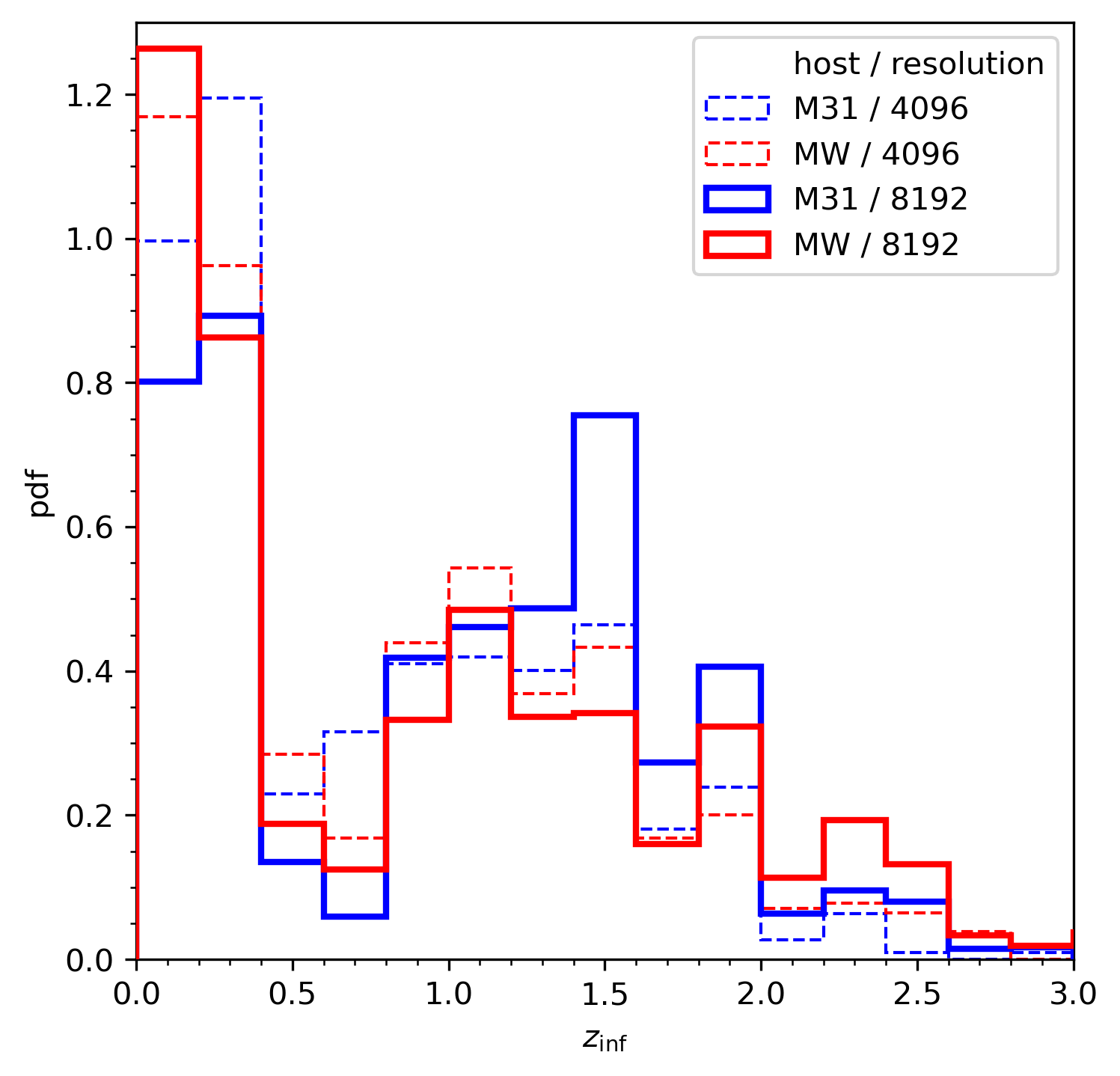}
\caption{Distribution of redshifts at infall $z_\mathrm{inf}$ of all satellites that went through $2 \times R_{200}$ of M31 (blue)  or the MW (red) as a probability density function. Thin dashed lines represent the simulation with a lower resolution (4096). The bold solid lines gives the results obtained from the higher resolution simulation (8192), i.e the bold lines in Figure \ref{fig:zhist}. In both cases, all realizations considered are merged.} 
\label{fig:4096_zhist}
\end{figure}

\begin{figure}
\centering
\includegraphics[width=0.4\textwidth]{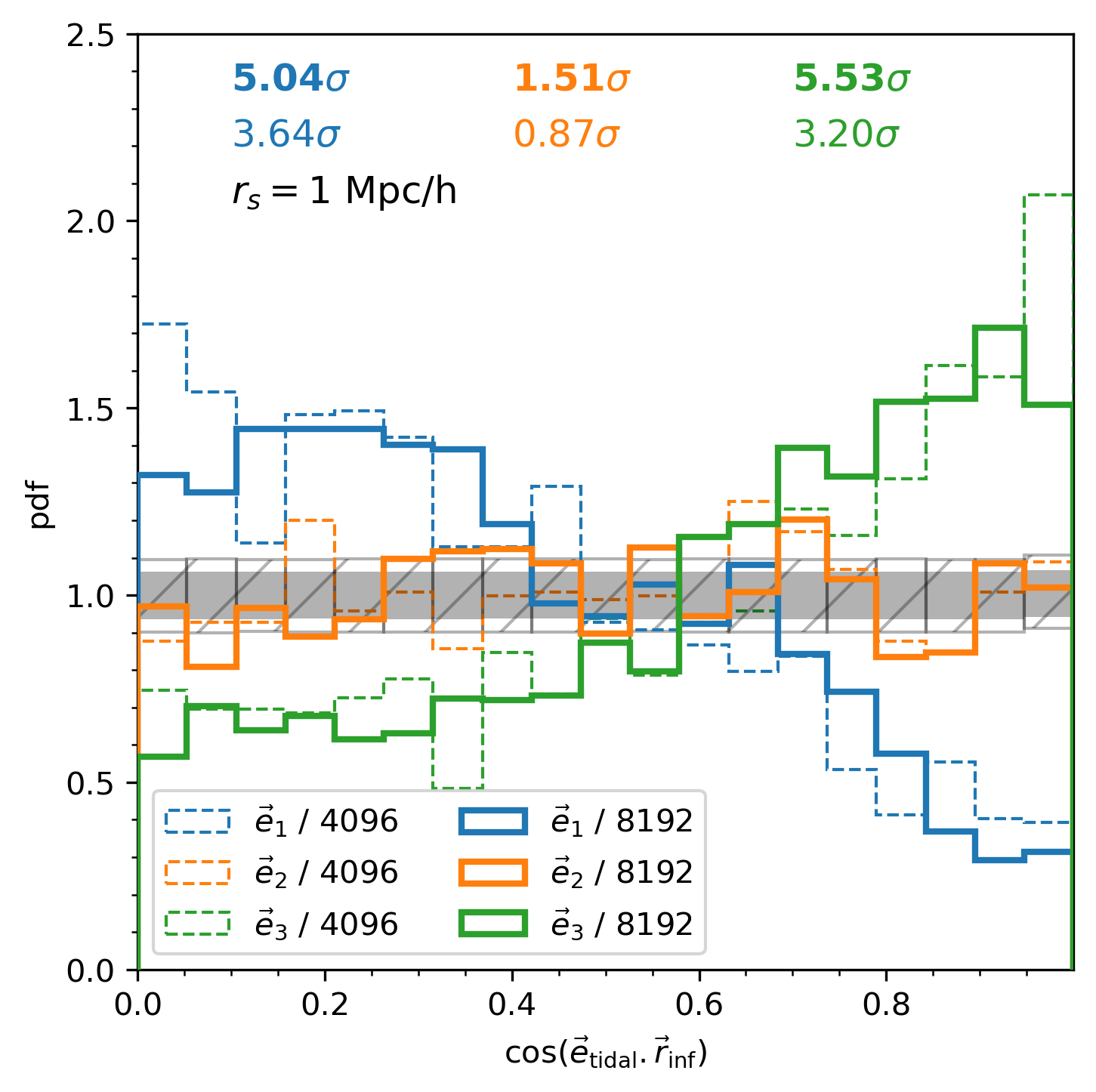}
\caption{Distribution of the angles between the infall direction $\vec{r}_\mathrm{inf}$ at $2 \times R_{200}$ and the three eigenvectors $\vec{e_1}$, $\vec{e_2}$ and $\vec{e_3}$ of the tidal tensor on which a smoothing of $r_s = 1$ Mpc has been applied. Distributions are quantified by means of probability density functions. The reader may refer to the caption of Figure \ref{fig:infalldir} for the description of the layout and the color code. Thin dashed lines represent the results given by the simulation with a lower resolution (4096), while the bold solid lines gives the results obtained from the higher resolution simulation (8192), i.e the bold lines in the bottom left panel of Figure \ref{fig:infalldir}. In both cases, all realizations considered are merged. For clarity purposes, only the regions corresponding to the $\pm 1 \sigma$ threshold are shown, where the hatched one corresponds to the lower resolution simulation, and the one filled in grey corresponds to the higher resolution simulation.}
\label{fig:4096_2rvir}
\end{figure}

The results obtained from the simulation with $4096^3$ particles are plotted in dashed lines in Figures \ref{fig:4096_zhist} and \ref{fig:4096_2rvir}, corresponding respectively to Figure \ref{fig:zhist} and the bottom left panel of Figure \ref{fig:infalldir}. They are compared to the results presented in this manuscript, represented in bold lines, and corresponding to the bold lines as well in both Figures \ref{fig:zhist} and \ref{fig:infalldir} (bottom left panel). 

In both figures, we observe similar patterns between the two resolutions. More importantly, Figure \ref{fig:4096_2rvir} shows that the alignment of the infall with the axis of slowest collapse $\vec{e_3}$, present in the higher resolution results, can still be noticed in the lower resolution simulation.

\bsp	
\label{lastpage}
\end{document}